\documentstyle[aps,12pt,epsf,amssymb]{revtex}
\begin{document}
\baselineskip=0.8cm
\newcommand{\ini}{\begin{equation}}
\newcommand{\fin}{\end{equation}}
\newcommand{\inir}{\begin{eqnarray}}
\newcommand{\finr}{\end{eqnarray}}
\newcommand{\inif}{\begin{figure}}
\newcommand{\finf}{\end{figure}}
\newcommand{\bc}{\begin{center}}
\newcommand{\ec}{\end{center}}
\def\ol{\overline}
\def\pa{\partial}
\def\ra{\rightarrow}
\def\ts{\times}
\def\df{\dotfill}
\def\bs{\backslash}
\def\dg{\dagger}

$~$

\hfill DSF-T-99/30

\vspace{1 cm}

\centerline{\LARGE{Neutrino masses and mixings}}

\centerline{\LARGE{in a seesaw framework}}

\vspace{1 cm}

\centerline{\large{D. Falcone}}

\vspace{1 cm}

\centerline{Dipartimento di Scienze Fisiche, Universit\`a di Napoli,}
\centerline{Mostra d'Oltremare, Pad. 19, I-80125, Napoli, Italy}

\centerline{{\tt e-mail: falcone@na.infn.it}}

\vspace{1 cm}

\begin{abstract}

\noindent
Assuming the seesaw mechanism for hierarchical neutrino masses, we calculate
the heavy neutrino masses under the hypotheses that the mixing in the Dirac
leptonic sector is similar to the quark mixing ($V_D \sim V_{CKM}$) and that
$M_{\nu} \sim M_u$ or $M_e$, where $M_{\nu}$ is the Dirac mass matrix of neutrinos.
As a result we find that for $M_{\nu} \sim M_u$ the vacuum oscillation solution
of the solar neutrino problem leads to a scale for the heavy neutrino mass well
above the unification scale, while for the MSW solutions there is agreement
with this scale. For $M_{\nu} \sim M_e$ the vacuum solution is consistent with
the unification scale, and the MSW solutions with an intermediate scale.
The mass of the lightest heavy neutrino can be as small as $10^5$ GeV.

\noindent
PACS numbers: 12.15.Ff, 14.60.Pq

\end{abstract}

\newpage

\section{Introduction}

\noindent
In the Minimal Standard Model (MSM)\cite{msm} the neutrino is massless.
This is because
with only the left-handed neutrino $\nu_L$ we cannot build a Dirac mass term,
and with only the Higgs doublet $\phi$ we cannot build a Majorana mass term
for $\nu_L$ after spontaneous symmetry breaking.
However, strong indications for nonzero neutrino mass come from
solar and atmospheric neutrino experiments and on the theoretical side
there are several extensions of the MSM that lead to nonzero neutrino mass
\cite{lan}.

The simplest one is to add the right-handed neutrino $\nu_R$ in order to have the
analogous of the quark $u_R$ in the leptonic sector. When this is done, it becomes
possible to give a Dirac mass to the neutrino by means of the same mechanism
used for the other fermions. Thus we expect this mass to be of the same order
of magnitude of the other fermion masses. Moreover, it is now also possible
to have
a bare Majorana mass term for $\nu_R$, and the corresponding value of the mass
is not constrained if the gauge group is the same of the MSM. Therefore
we have a new mass scale in the extended theory \cite{wil}, and it is a key
problem to understand if this new scale is associated to new physics,
that is a larger gauge group,
and at what energy it eventually happens.

If the Dirac mass of the neutrino is of the same order of the other quark or
lepton masses, the seesaw mechanism \cite{ss} relates the smallness of
the neutrino
mass to a very large scale in the Majorana term. Of course we have three
generations of fermions and we expect three light neutrinos and three heavy
ones. We assume that the light neutrino mass spectrum is hierarchical
as it happens for the quark and charged lepton mass spectra. We denote
by $m_1$, $m_2$, $m_3$ the Dirac masses, by $M_1$, $M_2$, $M_3$ the heavy
neutrino masses, and by $m_{\nu_1}$, $m_{\nu_2}$, $m_{\nu_3}$ the light
neutrino masses. From the solar and atmospheric neutrino experiments we can
infer \cite{bg}, with some uncertainty, the values of $m_{\nu_2}$ and $m_{\nu_3}$,
and of the neutrino mixing matrix $U$,
\ini
\nu_{\alpha}=U_{\alpha i} \nu_i,
\fin
where $U$ is unitary and connects the mass eigenstates $\nu_i$ ($i=1,2,3$) to
the weak eigenstates $\nu_{\alpha}$ ($\alpha=e,\mu,\tau$).

The aim of this paper is to calculate the heavy neutrino masses under
simple hypotheses on the Dirac masses of neutrinos and on the
matrix
\ini
V_D=V^{\dg}_{\nu} V_e
\fin
which is the analogous of the CKM matrix in the Dirac leptonic sector.
Namely, we assume $V_D \sim V_{CKM}$, and $M_{\nu} \sim M_u$
or $M_{\nu} \sim M_e$.
For $m_{\nu_1}$ we allow a variation of three orders in the hierarchical
regime. We are mostly interested in $M_3$
(the mass of the heaviest right-handed neutrino),
which is related to the
new mass scale of the theory, and in $M_1$
(the mass of the lightest right-handed neutrino), which has some importance in
baryogenesis $via$ leptogenesis \cite{bb,fy}. 
In Grand Unified Theories (GUTs)
$M_3$ is associated to unification or intermediate
scales \cite{dkp},
thus we match our results with these scales.
General considerations on heavy neutrino masses in the seesaw mechanism can be
found in ref.\cite{smir}. In the present paper we give a numerical analysis based
on the hypotheses above and the experimental data on solar and atmospheric
neutrinos.
                            
\section{Theory}

\noindent
Let us briefly explain the effect of the seesaw mechanism on the leptonic mixing.
The part of the Lagrangian we have to consider is
\ini
\ol{e}_L M_e e_R+\ol{\nu}_L M_{\nu} \nu_R+g \ol{\nu}_L e_L W+\ol{\nu}^c_L M'_R
\nu_R
\fin
where $M_e$ and $M_{\nu}$ are the Dirac mass matrices of charged leptons and
neutrinos respectively, and $M'_R$ is the Majorana mass matrix of right-handed
neutrinos.
If we assume the elements of $M'_R$ much greater than those of
$M_{\nu}$, the seesaw mechanism leads to the effective Lagrangian
\ini
\ol{e}_L M_e e_R+\ol{\nu}_L M'_L \nu^c_R+g \ol{\nu}_L e_L W+\ol{\nu}^c_L M'_R 
\nu_R
\fin
where
\ini
M'_L=M_{\nu} {M'_R}^{-1} M_{\nu}^T
\fin
is the Majorana mass matrix of left-handed neutrinos (in this context the
left-handed neutrinos are called light neutrinos and the right-handed neutrinos
are called heavy neutrinos).
Diagonalization of $M_e$, $M_L'$ gives (renaming the fermion fields)
\ini
\ol{e}_L D_e e_R+\ol{\nu}_L D_L \nu^c_R+g \ol{\nu}_L V_{lep} e_L W+
\ol{\nu}^c_L M'_R \nu_R.
\fin
Of course we can also diagonalize $M'_R$ without changing other parts of this
Lagrangian. The unitary matrix $V_{lep}$ \cite{mns} describes the weak
interactions
of light neutrinos with charged leptons. The following steps clarify the
structure of $V_{lep}$.

If in eqn.(3) we first diagonalize $M_e$ and $M_{\nu}$, obtaining
\ini
\ol{e}_L D_e e_R+\ol{\nu}_L D_{\nu} \nu_R+g \ol{\nu}_L V_D e_L W
+\ol{\nu}^c_L M_R \nu_R,
\fin
then the seesaw mechanism gives
\ini
\ol{e}_L D_e e_R+\ol{\nu}_L M_L {\nu}^c_R+g \ol{\nu}_L V_D e_L W
+\ol{\nu}^c_L M_R \nu_R
\fin
with
\ini
M_L=D_{\nu} M_R^{-1} D_{\nu}.
\fin  
Then, we diagonalize also $M_L$,
\ini
\ol{e}_L D_e e_R+\ol{\nu}_L D_L {\nu}^c_R+g \ol{\nu}_L V_s V_D e_L W+
\ol{\nu}^c_L M_R \nu_R,
\fin
and, comparing with eqn.(6), we recognize that
\ini
V_{lep}=V_s V_D
\fin
where
\ini
V_s M_L V_s^T=D_L.
\fin
We also understand that 
\ini
V_{lep}=U^{\dg},
\fin     
and point out that $M'_R$ ($M'_L$) differs from $M_R$ ($M_L$) by a
unitary transformation, hence they have the same eigenvalues.

In the Lagrangian (3) it is   possible to diagonalize $M_e$ or $M_R$
without changing the observables quantities. The same is not true
for $M_{\nu}$. Moreover, $M_e$ and $M_R$ can be diagonalized simultaneously.
In the Lagrangian (4) the following matrices can be diagonalized: $M_e$,
$M_L$, $M_R$, both $M_L$ and $M_R$, both $M_e$ and $M_R$. When we set $M_e=D_e$ in
eqn.(4) we have $M_L=U D_L U^T$, and when we set $M_L=D_L$ we get
$M_e=U^{\dg} D_e U$ if $M_e$ is chosen hermitian
or $M_e M_e^{\dg}=U^{\dg} D_e^2 U$ if $M_e$ contains three zeros \cite{falc}.

\section{Experiment}

\noindent
Experimental informations on neutrino masses and mixings are increasing
rapidly. To be definite we refer to \cite{bg}, where the matrix $U$ is
written as
\ini
U=\left(\begin{array}{ccc}
       c_{12} & s_{12} & 0 \\
     -s_{12}c_{23} & c_{12}c_{23} & s_{23} \\
      s_{12}s_{23}  & -c_{12}s_{23} & c_{23}
    \end{array}\right).
\fin
There is a zero in position 1-3, although it is only constrained to be much less
than one \cite{nrs}. 
The experimental data on oscillation of atmospheric and solar neutrinos lead
to three possible numerical forms for $U$,
corresponding to the three solutions of
the solar neutrino problem, namely small mixing and large mixing MSW
(smMSW and lmMSW) \cite{msw}, and vacuum oscillations (VO) \cite{mns,gp,vo}.
Choosing the central values of neutrino masses and of $s_{12}$ and $s_{23}$
from  ref.\cite{bg}, we have always  $m_{\nu_3}=5.7 \times 10^{-11}$ GeV, and
\ini
U = \left(\begin{array}{ccc}
       1 & 0.04 & 0 \\
     -0.032 & 0.80 & 0.60 \\
      0.024 & -0.60 & 0.80
    \end{array}\right) \equiv U_1,
\fin
$m_{\nu_2}=2.8 \times 10^{-12}$ GeV for small mixing MSW,
\ini
U = \left(\begin{array}{ccc}
      0.91  & 0.42 & 0 \\
     -0.336 & 0.726 & 0.60 \\
      0.252 & -0.544 & 0.80
    \end{array}\right) \equiv U_2,
\fin
$m_{\nu_2}=4.4 \times 10^{-12}$ GeV for large mixing MSW, and
\ini
U = \left(\begin{array}{ccc}
       0.80 & 0.60 & 0 \\
     -0.474 & 0.632 & 0.61 \\
      0.366 & -0.488 & 0.79
    \end{array}\right) \equiv U_3,
\fin
$m_{\nu_2}=9.2 \times 10^{-15}$ GeV for vacuum oscillations. 
We also consider maximal and bimaximal \cite{bim} mixing as limiting cases of
$U_1$ and $U_3$, respectively:
\ini
U_m = \left(\begin{array}{ccc}
       1 & 0 & 0 \\
      0 & \frac{1}{\sqrt2} & \frac{1}{\sqrt2} \\
      0 & -\frac{1}{\sqrt2} & \frac{1}{\sqrt2}
    \end{array}\right),
\fin
\ini
U_b = \left(\begin{array}{ccc}
       \frac{1}{\sqrt2} & \frac{1}{\sqrt2} & 0 \\
      -\frac{1}{2} & \frac{1}{2} & \frac{1}{\sqrt2} \\
       \frac{1}{2} & -\frac{1}{2} & \frac{1}{\sqrt2}
    \end{array}\right).
\fin
Experimental data on oscillations only give $\Delta m_{32}^2$,
$\Delta m_{21}^2$, from which, for hierarchical light neutrino masses, we yield
the values of $m_{\nu_3}$, $m_{\nu_2}$, because
$\Delta m_{32}^2 \simeq m_{\nu_3}^2$, $\Delta m_{21}^2 \simeq m_{\nu_2}^2$.
For $m_{\nu_1}$ we will assume $m_{\nu_1} \le 10^{-1} m_{\nu_2}$.
From the unitary matrices written above we see that leptonic mixing between
second and third family is large, while the mixing between first and second
family may be large or small. It is well-known that in the quark sector
all mixings are small.

\section{Calculation}

\noindent
Our determination of $M_1,~M_2,~M_3$ is based on the following assumptions.
Looking at eqn.(11) we see that $V_s$ could be responsible of the
enhancement of lepton mixing \cite{sm}. Therefore, it is suggestive 
to assume that the matrix $V_D$ has just  the form of the CKM matrix
\ini
V_D = \left(\begin{array}{ccc}
       1-\frac{1}{2}\lambda^2 & \lambda  & \lambda^4 \\
      -\lambda & 1-\frac{1}{2}\lambda^2 & \lambda^2  \\
     \lambda^3-\lambda^4 & -\lambda^2 & 1
    \end{array}\right).
\fin
This is similar to the form which might originate from GUTs \cite{bky}
\ini
V_D = \left(\begin{array}{ccc}
       1-\frac{1}{18}\lambda^2 & \frac{1}{3}\lambda  & \lambda^4 \\
      -\frac{1}{3}\lambda & 1-\frac{1}{18}\lambda^2 & \lambda^2  \\
    \frac{1}{3}\lambda^3-\lambda^4 & -\lambda^2 & 1
    \end{array}\right),
\fin
the difference being in the element $V_{D12}$, and this, in turn,
to the $V_D$ which results from the analogy of ref.\cite{fal}, where
$V_{D23} \simeq 2 \lambda^2$.
We also assume $M_{\nu} \sim M_u$ or $M_{\nu} \sim M_e$.
In particular,
\ini
M_{\nu} = D_{\nu} = \frac{m_{\tau}}{m_b} D_u,
\fin
where the factor is due to running \cite{acpr},
or
\ini
M_{\nu} = D_{\nu} = D_e.
\fin
We use quark (and charged lepton) masses at the scale $M_Z$ as in
ref.\cite{fk}.
It is important to notice that the values $M_1,M_2,M_3$
do not depend on the assumption $M_{\nu}=D_{\nu}$, because $M_R$ undergoes
a unitary transformation.
Also $V_D$ does not change if we rotate $e_L$ as $\nu_L$.
In fact, one can always diagonalize $M_u$ without changing  $V_{CKM}$
\cite{ma}, and $M_{\nu}$ without changing both $V_D$ and $M_1,M_2,M_3$.
We vary $m_{\nu_1}$ by three orders down from $m_{\nu_1}=10^{-1}m_{\nu_2}$
to $m_{\nu_1}=10^{-4}m_{\nu_2}$.
In the tables we report our results (notation: $xey \equiv x \times 10^y$, in
GeV).
They are obtained in the following way. From eqn.(11) we have 
\ini
V_s=V_{lep} V_D^{\dg};
\fin
using eqn.(12) we get
\ini
M_L=V_s^T D_L V_s,
\fin 
and from eqn.(9) we obtain
\ini
M_R=D_{\nu} M_L^{-1} D_{\nu}
\fin
and then its eigenvalues.
We see that in the case $M_{\nu} \sim M_u$ the VO solution leads to a scale
for $M_3$ well above the unification scale (around the Planck scale),
while the MSW solutions are
consistent with this scale. $M_2$ is around the intermediate scale.
In the case $M_{\nu} \sim M_e$
the VO solution gives $M_3$ near the unification scale,
while the MSW solutions bring it
near an intermediate scale. Also we notice that $M_1$ may be of the order $10^6$,
a relatively small value \cite{fal,bb}. The huge value of $M_3$ in the VO case
is due mainly to the lower values of $m_{\nu_1}$, $m_{\nu_2}$ respect to the
MSW case.
There is no substantial change if the factor
$m_{\tau}/m_b$ is erased from eqn.(22): the numerical
results are rescaled by the value 2.8.
We have introduced such a factor because the relation $M_u \sim M_{\nu}$
is typical of GUTs, where it is true at the unification scale, while the
factor $m_{\tau}/m_b$ appears at low energy due to running.
It can be checked that there is not an essential difference between the results
obtained by eqn.(20) ($V_D \sim V_{CKM}$) and those obtained by eqn.(21)
($V_D \sim V_{GUT}$). In fact, numbers differ by no more than one order of
magnitude.
Moreover, comparing values in the two MSW cases, the
effect of changes in $m_{\nu_1}$ is appearent: in the small mixing
solution $M_3$ varies by one order, in the large mixing solution by three
orders. If we want $M_3$ not to exceed the unification scale, then $m_{\nu_1}$
cannot be much smaller than $m_{\nu_2}$, in the large mixing MSW.
Maximal and bimaximal mixings confirm the results obtained for small mixing MSW
and vacuum oscillations, respectively.

\newpage

\begin{center}
\begin{tabular}{|c|ccccc|}\hline
$~$ & smMSW & lmMSW & VO & max & bimax \\ \hline
$M_1$ & 1.3e6 & 6.8e5 & 2.5e6 & 1.1e6 & 4.0e6 \\
$~$ & 1.6e6 & 2.3e6 & 3.1e9 & 1.2e6 & 3.7e9 \\ \hline
$M_2$ & 3.1e10 & 1.5e10 & 1.7e12 & 3.0e10 & 1.4e12 \\
$~$ & 9.4e11 & 2.0e10 & 2.0e12 & 1.1e13 & 1.5e12 \\ \hline
$M_3$ & 1.7e15 & 2.8e15 & 2.2e18 & 2.2e15 & 3.7e18 \\
$~$ & 4.7e16 & 1.9e18 & 1.9e21 & 5.0e15 & 3.3e21 \\ \hline
\end{tabular}
\end{center}

\centerline{Table 1: $V_D \sim V_{CKM},M_{\nu} \sim M_u$}
$~$

\begin{center}
\begin{tabular}{|c|ccccc|}\hline
$~$ & smMSW & lmMSW & VO & max & bimax \\ \hline
$M_1$ & 1.7e5 & 8.6e4 & 2.2e5 & 1.4e5 & 1.7e5 \\
$~$ & 2.0e5 & 9.7e4 & 1.3e7 & 1.6e5 & 1.4e7 \\ \hline
$M_2$ & 2.1e9 & 1.0e9 & 1.2e11 & 2.0e9 & 9.3e10 \\
$~$ & 5.4e10 & 1.3e9 & 1.4e11 & 3.8e11 & 1.0e11 \\ \hline
$M_3$ & 5.0e11 & 7.9e11 & 6.1e14 & 6.2e11 & 1.0e15 \\
$~$ & 1.5e13 & 5.3e14 & 5.2e17 & 2.8e12 & 9.3e17 \\ \hline
\end{tabular}
\end{center}

\centerline{Table 2: $V_D \sim V_{CKM},M_{\nu} \sim M_e$}
$~$

\newpage

As a matter of fact $V_D \sim V_{CKM}$ and $V_D \sim V_{GUT}$ are not so
different from $V_D \sim I$. In such a case $V_{lep} \simeq V_s$. The
opposite case is $V_{lep} \simeq V_D$ and then $V_s \simeq I$, that is, when
$M_{\nu}=D_{\nu}$ also $M_R=D_R$. From the seesaw mechanism we obtain
\ini
M_i = \frac{m_i^2}{m_{\nu_i}},
\fin
which gives
$M_3 \sim 10^{14}$,  $M_2 \sim 10^{10}$,
$M_1 \sim 10^6 \div 10^{9}$ (MSW), $M_2 \sim 10^{13}$,
$M_1 \sim 10^9 \div 10^{12}$ (VO) GeV in the
case $M_{\nu} \sim M_u$;
$M_3 \sim 10^{10}$,  $M_2 \sim 10^{9}$,
$M_1 \sim 10^5 \div 10^{8}$ (MSW), $M_2 \sim 10^{12}$,
$M_1 \sim 10^8 \div 10^{11}$ (VO) GeV in the
case $M_{\nu} \sim M_e$. In the VO solution with $M_{\nu} \sim M_e$,
$M_2$ exceeds $M_3$, and also $M_1$ can do it.

Let us now briefly discuss the sensitivity to input mixing angles of the
results reported in the first three columns of tables 1,2. By allowing
$s_{12}$ and $s_{23}$ to vary inside the ranges reported in ref.\cite{bg}
we have found that the numerical values of right-handed neutrino masses
change by no more than one order of magnitude. The same happens if
$U_{13}$ is different from zero up to 0.1. Therefore the above considerations
on the physical scales do not change.

It is also interesting to match our results, obtained by a hierarchical
spectrum, with the degenerate spectrum and the democratic mixing \cite{tre}
\ini
U_d = \left(\begin{array}{ccc}
       \frac{1}{\sqrt2} & \frac{1}{\sqrt2} & 0 \\
      -\frac{1}{\sqrt6} & \frac{1}{\sqrt6} & \frac{2}{\sqrt6} \\
       \frac{1}{\sqrt3} & -\frac{1}{\sqrt3} & \frac{1}{\sqrt3}
    \end{array}\right).
\fin
Assuming as light neutrino mass $m_0=2$ eV, relevant for hot dark matter,
and the democrating mixing $U=U_d$, we obtain
$M_1=9.2 \ts 10^{2}$, $M_2=7.7 \ts 10^{7}$, $M_3=5.5 \ts 10^{12}$ GeV
for $M_{\nu} \sim M_u$ and
$M_1=1.2 \ts 10^{2}$, $M_2=5.3 \ts 10^{6}$, $M_3=1.5 \ts 10^{9}$ GeV
for $M_{\nu} \sim M_e$,
that is $M_3$ at the intermediate scale and $M_1$ even at the electroweak scale.
From eqns.(25),(26) we see that in the
case of degenerate masses $M_R$ is proportional to $1/m_0$ and one can easily
obtain $M_3$ when $m_0$ is lowered.

\section{Conclusion}

\noindent
We have calculated the heavy neutrino masses in a seesaw framework, under
simple hypotheses on the Dirac sector, and using experimental limits on
light neutrino masses and mixings.
The results have been matched with intermediate and unification
scales. A key result is that the large mixing MSW solution can be reconciled
with GUTs.
The analysis can be improved when more
precise data will be available. Also the effect of phases should be
considered \cite{tan}. There are several recent studies about
the seesaw mechanism \cite{see}, based on various forms of mass matrices;
a nice review is in ref.\cite{lo}.
Instead, in this paper,
we work on the matrix $V_D$ and on $D_{\nu}$,
that is the leptonic quantities which correspond, in the quark sector, to the
observable quantities.

\end{document}